\documentclass[11pt,a4paper]{article}
\usepackage[T1]{fontenc}
\usepackage[utf8]{inputenc}
\usepackage{amsmath,amssymb}
\usepackage{graphicx}
\usepackage{xcolor}
\usepackage[hidelinks]{hyperref}
\usepackage{natbib}
\usepackage{geometry}
\usepackage{tikz}
\usepackage{microtype}
\usepackage{booktabs}
\usetikzlibrary{positioning,arrows.meta}
\providecommand{\keywords}[1]{\vspace{0.5em}\noindent\textbf{Keywords:} #1\vspace{1em}}

\title{\Large \textbf{Empathy as Predictive Misalignment Tolerance:\\
A Co-Regulation Framework and the\\
Regime Structure of Dialogue Repair}}
\author{Molood ARMAN \\
Independent Researcher \\
\texttt{arman.molood@gmail.com}}
\date{}

\begin{document}
\maketitle

\begin{center}
\textit{To all forms of intelligence learning to understand one another.}
\end{center}

\vspace{1em}

\begin{abstract}
\noindent Empathy is most often theorised as resonance: a mirroring or simulation of another's present emotional or cognitive state. This synchronic framing has shaped how empathy is operationalised in artificial systems, where ``empathic'' behaviour is typically defined as affect recognition and response alignment. We argue that this is the wrong target for systems engaged in extended dialogue. Genuine understanding unfolds over time, shaped by prediction, divergence, and repair.

We propose a reframing of empathy as \textbf{\textit{predictive misalignment tolerance}}: the capacity to anticipate how interpretations will diverge across time and to regulate that divergence within a bounded interpretive space rather than collapse it. We formalise this intuition as \textbf{Interpretive Error Tolerance (IET)}, a dynamic-threshold heuristic that models empathy as the maintenance of a viable band of divergence between interacting agents. We position IET as a regulatory \textit{lens}, not a validated controller --- a way of asking what kind of alignment we want artificial systems to cultivate.

We then report results from two computational probes designed to operationalise this framework under controlled noise conditions. The IET update rule, as currently specified, does not outperform fixed-threshold baselines on either probe. The experiments instead yield a distinct and robust empirical finding: dialogue repair exhibits a \textit{regime-dependent structure}, trading off discriminative fidelity for gist preservation. Repair mechanisms degrade retrieval accuracy at low noise levels but begin to preserve gist meaning under high-noise conditions, revealing a previously under-characterised interaction between noise level, repair action, and evaluation metric.

We interpret this regime structure through the lens of IET, arguing that empathy in extended interaction is not the elimination of divergence but the regulation of its dynamics across time. This perspective suggests a shift in the design of empathic AI systems --- from convergence-seeking behaviour toward the explicit management of interpretive distance.
\end{abstract}


\section{From Resonance to Anticipatory Understanding}

Classical accounts of empathy in psychology and cognitive neuroscience emphasise either affective resonance --- a felt mirroring of another's present state --- or cognitive simulation, in which an observer reconstructs another's perspective by mental projection \citep{decety2004functional,zaki2014empathy}. Both framings are essentially synchronic: empathy is what happens \textit{in the moment} of contact, and successful empathy is convergence in that moment.

This synchronic picture has migrated, often uncritically, into the design of empathic artificial systems. Conversational agents are described as ``empathic'' when they recognise emotion in user input and produce affectively congruent output \citep{sharma2023human,shen2024empathy}. This is not nothing --- recognising distress in a help-seeking user matters --- but it is a thin notion of empathy, one that mistakes affective imitation for understanding. It cannot account for what happens when a conversation extends across many turns, when topics shift, when one party's frame of reference moves while the other's does not, when something said five minutes ago retroactively changes the meaning of what is being said now.

In ordinary human conversation, understanding is not a state attained and held; it is a temporal process that drifts, fractures, and is repaired \citep{clark1991grounding,garrod2004conversation,pickering2021understanding}. Misunderstanding is not the failure of empathy but one of its working materials. We therefore propose a different framing: empathy is the capacity to \textit{predict} how an interlocutor's interpretation will diverge from one's own across time, and to regulate that divergence proactively. On this view, empathy is fundamentally a temporal regulatory process. It is anticipatory rather than reactive, and its measure is not the depth of momentary alignment but the sustained ability to absorb, repair, and learn from divergence.

This paper develops that framing into a coherent conceptual position. We call the central proposal \textbf{Interpretive Error Tolerance (IET)}: empathy as the regulation, not the elimination, of misalignment. Three commitments organise the argument. First, empathy in extended interaction is constituted by \textit{tolerance of bounded divergence}, not by convergence. Second, this tolerance is not a fixed trait but a \textit{dynamic threshold} that interlocutors continuously co-adapt. Third, the design of empathic artificial systems should reflect this: it should aim at \textit{co-regulation} of meaning rather than at correctness or affective mimicry.

Crucially, our empirical results do not validate IET as a superior control mechanism in its current form. Instead, they reveal a regime-structured behaviour of dialogue repair that existing evaluation frameworks tend to obscure. We therefore position IET not as a validated controller, but as a conceptual lens through which this regime structure can be understood and further investigated. The contribution of this paper is conceptual, normative, and methodological: we integrate strands from predictive processing, enactive cognition, phenomenology, and the ethics of alterity; draw out their joint implications for human--AI design; and report findings from two falsifiable computational probes that, while not validating IET as a controller, surface a structural feature of dialogue repair the field has under-engaged.

\section{Theoretical Foundations}

\subsection{Predictive Processing and the Sociality of Inference}
Predictive-processing frameworks model cognition as hierarchical inference under prediction-error minimisation \citep{friston2010free,clark2013whatever,hohwy2013predictive}. The brain, on this view, is not a passive receiver of sensation but a generative system that continuously forecasts incoming evidence and updates its internal models when forecasts fail. Most of the foundational literature concerns perception and action in a single embodied agent, but extensions to social cognition have been developed in two main directions: theory-of-mind models in which one agent maintains a generative model of another's hidden states \citep{rabinowitz2018machine}, and dialectical or coupling accounts in which two predictive systems mutually condition each other's inferences \citep{pickering2013integrated,bolis2018observing,pickering2021understanding}.

\citet{pickering2021understanding} argue that successful dialogue depends on the active matching of \textit{forward predictions} between interlocutors --- each anticipating the other's next contribution and updating their own production accordingly. \citet{bolis2018observing} go further, framing many social-cognitive difficulties not as deficits in any individual but as ``dialectical misattunement'': failures of inter-agent predictive coupling that emerge in the dynamic relation rather than in either agent in isolation. These are conceptual neighbours of the IET framing. The shared insight is that the interesting variable in dialogue is not what either agent represents in isolation, but the \textit{distance} and \textit{trajectory} between two evolving interpretive systems.

What IET adds to this picture is a precise commitment about the regulatory target. The aim is not to minimise the distance between two interpretive trajectories --- that would collapse the difference between them. The aim is to keep the distance within a viable band. We turn to that commitment in Section~3.

\subsection{Co-Regulation, Embodiment, and Enactive Sense-Making}
Affective neuroscience has long described human interaction as a coupled dynamical system in which emotional, physiological, and behavioural signals are mutually adjusted to maintain interpersonal homeostasis \citep{feldman2017neurobiology,butler2013emotional}. Within enactive and embodied traditions, this insight is pushed further: meaning itself is not transferred between minds but is jointly produced through participatory sense-making \citep{colombetti2014feeling,di2018linguistic}. Empathy on the enactive view is not the recovery of another's state but a coordinated process of generating shared significance, in which neither participant's internal representation is the privileged reference.

For our purposes, two features of this literature matter. First, the relevant homeostasis is interpersonal, not intrapersonal: equilibrium is sustained \textit{between} agents, not within either of them alone. Second, this between-agent equilibrium is compatible with --- in fact requires --- continued internal difference. Co-regulation does not erase the distinction between participants; it modulates it. IET takes this lesson and asks: what is the analogous equilibrium for \textit{interpretive} content? What does it mean for two minds to remain in sustained communicative contact while continuing to interpret the world from genuinely distinct standpoints?

\subsection{Phenomenology: Co-presence Without Fusion}
Phenomenological accounts of intersubjectivity converge on a similar insight from a different direction. \citet{zahavi2025phenomenology} argues that the proper structure of empathy is co-presence rather than fusion: the experiencing of another \textit{as} another, not the absorption of their experience into one's own. The Levinasian tradition radicalises this point: the Other is constitutively beyond what I can grasp, and the ethical relation to the Other depends on that very irreducibility \citep{levinas1979totality}. To attempt total understanding is, on this view, already an ethical mistake --- a kind of conceptual violence performed in the name of empathy.

This matters for AI design in a specific way. If we frame the goal of empathic systems as ``understanding the user as well as possible,'' we are tacitly committed to a fusion model: more alignment is always better. If we instead frame the goal as \textit{sustaining a workable interpretive relationship while preserving the user's irreducibility}, the design problem changes. We are no longer optimising convergence; we are regulating distance. \citet{shukla2026relationship} make a parallel argument from the side of clinical practice: AI systems in mental-health settings should function as bridges that strengthen a user's wider relational ecology, not as substitutes that absorb relational space.

\subsection{The Gap}
Existing computational accounts of empathic AI sit largely within the synchronic, recognition-and-mimicry frame: detect affect, respond congruently, repeat \citep{sharma2023human,shen2024empathy}. Where temporal modelling enters, it usually does so as adaptation \textit{toward} greater alignment over time. Recent work in multi-agent LLM systems has begun to push back: \citet{wu2025hidden} argue that \textit{partial} alignment --- in which agents discuss but act on their own subjective interpretations --- outperforms forced consensus in volatile environments, especially under high uncertainty. This is conceptually adjacent to IET, but framed as a property of multi-agent task performance rather than as a theory of dyadic understanding. Critical work in AI ethics \citep{srinivasan2022role,kerasidou2020artificial} has separately argued that ``empathic'' systems can entrench rather than respect difference. What is missing is a positive account that synthesises these strands: a framework that takes temporality, divergence, and the ethical preservation of otherness as constitutive of empathy itself, rather than as nuisances to be minimised. The remainder of this paper develops one such account and reports the methodological findings of attempting to operationalise it.

\section{The IET Lens}

\subsection{Empathy as Bounded Misalignment Tolerance}
The core proposal is straightforward to state and consequential when taken seriously. Empathy in extended interaction is not the minimisation of interpretive divergence between agents; it is the maintenance of that divergence within a bounded, dynamically negotiated zone. We will refer to this zone as the agents' \textit{tolerance space}. A successful empathic exchange is one in which interpretive divergence remains inside the tolerance space across the conversation; an unsuccessful one is either an over-coupled exchange, in which the tolerance space collapses and one party's frame absorbs the other's, or an under-coupled exchange, in which divergence escapes any tolerance band and contact dissolves. Both failure modes are interesting. Both have analogues in well-known pathologies of human and human--machine interaction: enmeshment and pseudo-mutuality on one side, talking past each other on the other.

\subsection{A Formal Heuristic}
We can formalise this as a regulatory heuristic. Let $\Delta_t$ denote the interpretive divergence between two agents at conversational turn $t$, and $\theta_t$ the agents' current tolerance threshold. The IET update rule is:
\begin{equation}
\theta_{t+1} = \theta_t + \eta \,\bigl(\sigma(\Delta_t) - \sigma^{*}\bigr),
\label{eq:iet}
\end{equation}
where $\sigma(\Delta_t)$ is a short-window estimate of the variance of recent divergence, $\sigma^{*}$ is a homeostatic target for that variance, and $\eta$ is an adaptivity rate. When the variance of divergence rises above target, the tolerance threshold widens; when divergence becomes more orderly, the threshold tightens. A repair action --- a clarification, a rephrase, a slowing of the conversational pace, or, in the computational implementations explored in Section~\ref{sec:methodnote}, a downweighting of the suspect contribution to the running interpretive context --- is initiated when projected divergence threatens to leave the tolerance space.

We want to be candid about the status of equation~\eqref{eq:iet}. It is a \textit{conceptual heuristic}, not a validated control algorithm. We use it to give the framing precise structure --- to make explicit that interpretive homeostasis involves a \textit{variance-based} update rather than a level-based one, and that the controlled quantity is divergence around a target band rather than divergence \textit{tout court}. Whether this particular update rule, applied to a particular operationalisation of $\Delta_t$, improves any specific downstream task is a separate, empirical question. Section~\ref{sec:methodnote} reports findings from two attempts to answer it.

\subsection{Interpretive Homeostasis}
The state IET regulates is what we call \textit{interpretive homeostasis}: a sustained equilibrium in which two agents' interpretive trajectories remain related --- close enough to permit communication, distant enough to preserve the difference between them. The biological metaphor matters. Physiological homeostasis is not the absence of change; it is the regulation of change around viable set-points. Body temperature fluctuates constantly; what matters is that it stays in a band that supports the organism. Interpretive homeostasis is the analogous claim for meaning. Two minds in sustained empathic contact are not converging on a single interpretation; they are jointly maintaining a band of viable mutual intelligibility, within which their interpretations can vary, drift, and recover.

\begin{figure}[h!]
\centering
\begin{tikzpicture}[>={Stealth[length=2.5mm]},
                    every node/.style={font=\sffamily\small, align=center},
                    box/.style={draw, rounded corners=4pt, minimum width=2.8cm, minimum height=0.9cm, fill=gray!8}]
\node[box] (A) at (0,0) {Agent A};
\node[box] (B) at (8,0) {Agent B};
\node[box, fill=blue!8] (delta) at (4, 0.9) {Divergence $\Delta_t$};
\node[box, fill=green!8] (theta) at (4, -0.9) {Tolerance $\theta_t$\\(adaptive band)};
\draw[->] (A.north east) to[bend left=20] (delta.west);
\draw[->] (B.north west) to[bend right=20] (delta.east);
\draw[->] (delta) -- (theta) node[midway, right, font=\footnotesize]{variance update};
\draw[->] (theta.west) to[bend left=20] (A.south east);
\draw[->] (theta.east) to[bend right=20] (B.south west);
\end{tikzpicture}
\caption{Interpretive Error Tolerance as a bidirectional regulatory loop. Agents do not minimise divergence; they jointly modulate the tolerance band within which divergence is permitted to evolve.}
\label{fig:iet}
\end{figure}
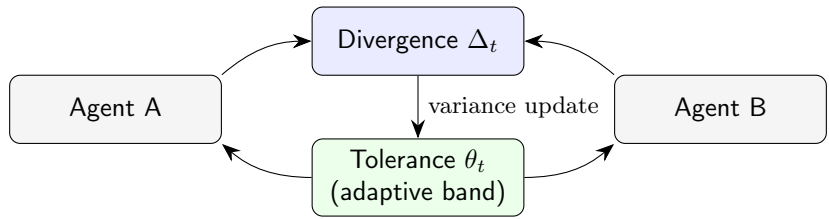

\section{Design Implications for Empathic AI}

If empathy is misalignment tolerance rather than alignment, what follows for the design of artificial conversational systems? We articulate four implications. Each is intentionally framed as a stance, not a recipe; the value of the IET lens is in reorienting design questions, not in dictating particular architectures.

\paragraph{Manage divergence; do not erase it.}
The dominant design pattern for empathic AI is convergence-seeking: the system reads user state and adjusts toward affective or interpretive alignment. The IET framing inverts the default. The design question becomes: what is the appropriate band of divergence for this user, this context, this stage of interaction? A too-narrow band produces an agent that smothers the user's distinctness, finishing their sentences with the system's framing. A too-wide band produces an agent that drifts, missing the points at which the user's interpretive trajectory has actually crossed into incoherence. Healthy empathic interaction requires neither extreme. It requires a workable tolerance band, and that band should be visible, contestable, and adjustable.

\paragraph{Empathy is regulation, not mimicry.}
Affect-mirroring is computationally cheap and rhetorically attractive, but it is a fragile foundation for sustained empathic interaction. A system that produces apparently empathic outputs by matching user affect can perform empathy without doing any of the regulatory work that empathy actually consists in. Worse, in extended interactions affect-mirroring may amplify rather than absorb interpretive drift: the system's mirrored output becomes the next turn's prior, and small misalignments compound. The IET lens reframes the design target. What we want from an empathic system is not that it look empathic at any given turn, but that it sustain a workable interpretive relationship across many turns --- which means active regulation, not surface congruence.

\paragraph{Tolerance must be personalised and contestable.}
Different users, different contexts, and different cultural settings call for different tolerance bands. A user in cognitive distress may need a tighter band, with more frequent clarification. A user in creative exploration may need a much wider band, in which the system's job is to stay engaged across genuinely surprising shifts of frame. A clinical setting calls for different defaults than a casual one. In each case, the parameters $\eta$ and $\sigma^{*}$ in equation~\eqref{eq:iet} are not given by the data; they are design choices that encode a normative stance about how much divergence is appropriate. Making those choices visible --- exposing them as user-adjustable settings rather than burying them in implicit defaults --- is itself an ethical commitment.

\paragraph{Misunderstanding is feedback, not failure.}
The recognition-and-mimicry view treats misunderstanding as a defect: the user's state was misread, and the system should do better. The IET view treats misunderstanding as the principal source of information about the state of the interpretive coupling. A system that never registers misunderstanding has either an excessively wide tolerance band (in which it fails to notice meaningful drift) or an excessively narrow one (in which it has homogenised the interaction). Either is a sign of unhealthy regulation. Designing systems that surface and learn from misunderstanding, rather than suppressing it, is more demanding than designing systems that perform fluent agreement, but it produces a different and arguably better object: an interlocutor that knows when it is at the edge of its understanding.

\section{Ethical and Societal Dimensions}

The IET framing is not ethically neutral, and we want to be explicit about its commitments and the questions they open.

\paragraph{Who controls $\theta_t$?}
Equation~\eqref{eq:iet} defines a control loop, and every control loop has a controller. In an empathic AI system the answer to ``who controls the tolerance threshold?'' is not given by the formalism; it is a design and governance choice. If $\theta_t$ is set by the platform operator and hidden from the user, the system's interpretive accommodation becomes a tool of soft power: the platform decides how much of the user's distinctness will be preserved. If $\theta_t$ is set by the user, the system becomes more like a mirror with adjustable curvature, but the user's ability to assess what setting is appropriate may itself be limited. If $\theta_t$ is jointly negotiated --- through visible system behaviour and explicit user preferences --- the system enters into something more like a relationship, with all the benefits and demands that implies.

We do not think there is one correct answer, but we think there is a wrong one: leaving the tolerance threshold implicit and unaccountable. An empathic system whose regulation of divergence is invisible to its user cannot be ethically evaluated, and the appearance of empathy becomes a mechanism for the very assimilation that empathy, properly understood, is supposed to resist.

\paragraph{Cultural and contextual variation.}
What counts as a comfortable interpretive band varies across cultures, generations, and communicative contexts. Norms about appropriate directness, about the role of clarification, about how much misunderstanding is acceptable before it becomes rude or threatening, differ widely. A system trained or tuned on one population's norms will impose those norms on others. The IET framing makes this explicit rather than papering over it: $\sigma^{*}$ is not universal, and any claim that it is is itself a cultural commitment. Empathic AI design should therefore include explicit reasoning about whose interpretive norms are being encoded, who participates in setting them, and how the system handles users whose norms differ from its defaults.

\paragraph{Alignment as control, alignment as co-adaptation.}
Contemporary AI alignment discourse has been shaped by a control-theoretic framing: the system is to be aligned with human values, and alignment is something done \textit{to} the system. The IET framing suggests a complementary picture for the dialogic case. In sustained interaction between humans and AI systems, alignment is not unilateral; it is a process in which both parties' interpretive states co-evolve. This does not dissolve the asymmetries --- the system's parameters, training data, and deployment context are set by humans, and rightly so --- but it does push back against the picture in which the user is a fixed target the system must hit. What the user means at turn 50 of an interaction is not what they meant at turn 1, partly because of how the system has responded in between. Ethical design must take this seriously: the system shapes the user's interpretive trajectory even as it tries to follow it, and a framing that pretends otherwise will systematically misdescribe what is happening.

\paragraph{The risk of empathy-as-capture.}
A final concern. If a system becomes very good at interpretive co-regulation, the very tolerance that makes it feel understanding can become a mechanism of dependency. Users may come to prefer interaction with the system over interaction with humans precisely because the system regulates divergence so smoothly. We do not think this is an argument against IET-style empathic design, but it is an argument for designing such systems with epistemic and relational humility: surfacing their limits, encouraging users to maintain human relationships, refusing to occupy interpretive space that the user would be better served by sharing with another person. \citet{shukla2026relationship} have argued in the clinical context that AI ought to function as a bridge into wider relational ecologies rather than as a substitute for them; we read this as the operational form of the same ethical commitment. A well-designed empathic system is one that helps the user stay in meaning, not one that makes itself the only place they can.

\section{What Two Empirical Probes Taught Us}
\label{sec:methodnote}

A natural question, raised by any conceptual proposal of this kind, is whether it can be empirically validated. We attempted two computational probes and report the findings here, because they are themselves part of the argument. We want to be transparent about what the probes tested, what they found, and what we take them to license.

\subsection{Setup}

Both probes used the same general structure. We sampled multi-turn dialogues from the DailyDialog corpus \citep{li2017dailydialog}, encoded each turn with a sentence-transformer model, and evaluated three repair policies in a controlled noise-injection paradigm: a \textbf{No-Repair} baseline, a \textbf{Fixed-$\theta$} repair policy, and an \textbf{Adaptive-IET} policy implementing equation~\eqref{eq:iet}. Time-varying noise was injected into the middle of the dialogue context (so that fixed-$\theta$ controllers could not optimally serve all phases simultaneously). All policies received the same drift signal; they differed only in how they used it. Critically, we evaluated on \textit{external} metrics --- next-utterance retrieval accuracy in a 1-of-N pool, and (in the second probe) cosine similarity to a clean-original reference context --- so that the controllers' behaviour could not directly optimise the score. The full notebooks and the pre-committed hyperparameter values are in the supplementary material; the design discipline we adopted across both probes was that hyperparameters were locked before each experiment was run, and the result was accepted regardless of which way it pointed.

\subsection{Probe 1: The Signal Choice Problem}

The first probe used the most direct operationalisation of $\Delta_t$ available: the cosine distance between each turn's embedding and the running summary of the surviving context. Repair was implemented as a hard drop of suspect turns. Across 5 noise rates and 200 dialogues per condition, the result was clear and informative: \textbf{both repair policies under-performed the no-repair baseline at every noise rate}, and adaptive and fixed were statistically indistinguishable from each other (paired McNemar $p=1.0$ at every rate; 0--1 discordant pairs out of 147 dialogues per condition).

The mechanistic explanation is the substantive finding. Mean turn-to-turn drift in \textit{clean} DailyDialog dialogues was already in the range $0.6\text{--}0.8$ cosine distance, because real conversation routinely introduces sub-topics, follow-up questions, and perspective shifts. With any reasonable threshold, the controller fired on most turns whether or not noise was present. At noise rate $0$, the precision of the repair decision was zero: every flagged turn was a false positive. By the time the controllers had finished ``repairing,'' they had discarded most of the conversational material, leaving the no-repair baseline with strictly more information from which to retrieve.

The lesson we took from this is not that embedding-based signals cannot detect misalignment. It is that \textit{the reference frame for measuring drift fundamentally determines what the signal carries}. Distance from a running context summary conflates two phenomena --- topical evolution and noise-induced misalignment --- that ought to be kept distinct.

\subsection{Probe 2: Anchor-Based Drift and Continuous Downweighting}

The second probe addressed this directly. We redesigned three components of the experiment. (i)~The drift signal was computed as the cosine distance between each turn's embedding and a \textit{topic anchor} formed from the first two turns of the dialogue (which were never noised). (ii)~The repair action was changed from a hard drop to continuous downweighting: a turn's contribution to the running context is multiplied by $w = \exp(-k \cdot \max(0, \Delta_t - \theta_t))$, so that turns near the threshold are softly attenuated rather than discarded. (iii)~We added a secondary metric --- cosine similarity between the controller's final context embedding and the clean-original (unnoised) context --- so that we could distinguish two notions of preservation: \textit{discriminative} (retrieval) and \textit{gist} (mean-aligned content).

\begin{figure}[h!]
\centering
\includegraphics[width=0.7\textwidth]{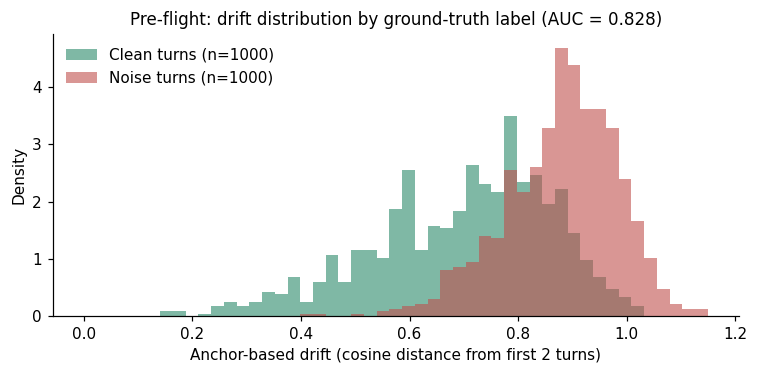}
\caption{Pre-flight check from Probe 2: distribution of anchor-based drift on clean vs. noise-injected turns. AUC $=0.828$. The redesigned signal cleanly separates the two regimes that Probe 1's running-summary signal had conflated.}
\label{fig:preflight}
\end{figure}

Before running the controller comparison, we ran a \textit{pre-flight check} on the redesigned signal: a simple ROC analysis testing whether anchor-based drift could distinguish ground-truth clean turns from injected noise turns. The AUC was $0.828$ (clean turns: drift $0.698 \pm 0.165$; noise turns: drift $0.879 \pm 0.103$; see Figure~\ref{fig:preflight}). The redesigned signal is informative; the failure of Probe 1 was a signal-choice failure, not a fundamental limit of embedding-based methods.

\subsection{What Probe 2 Found}

The headline result on the primary metric (1-of-20 retrieval accuracy) was again that adaptive and fixed policies were statistically indistinguishable from each other (paired McNemar $p=1.0$ at every noise rate; differences in mean accuracy of $0.000\text{--}0.005$ across 200 dialogues per condition). On the secondary metric (cosine to clean-original), paired-$t$ tests reported very small but nominally significant differences in adaptive's favour (mean differences $\approx 0.0001\text{--}0.0011$). With $n=200$ paired observations, sub-millisigma differences become statistically detectable; we do not claim them as substantively meaningful.

We take this as a clean negative result on the IET update rule \textit{as currently formalised}: under the parameter values we pre-committed to, the variance signal $\sigma(\Delta_t)$ moves $\theta_t$ across only a small range (in our run, from $0.500$ to $0.508$ over seven turns), and that movement is too small relative to the action range to differentiate adaptive from fixed behaviour. Whether a more aggressive adaptivity rate $\eta$ would change this is an interesting question, but one we declined to pursue because doing so would amount to tuning until adaptive wins --- precisely the kind of post-hoc rescue this paper has tried to avoid.

\subsection{The Substantive Finding: Repair is Regime-Structured}

A second, unexpected finding emerged from the dual-metric design, and we think it is the more important contribution of the empirical work. The two metrics tell substantially different stories about whether repair helps at all. Table~\ref{tab:regime} summarises the result.

\begin{table}[h!]
\centering
\small
\begin{tabular}{lccccc}
\toprule
& \multicolumn{2}{c}{Primary: retrieval accuracy} & \multicolumn{2}{c}{Secondary: cos-to-clean} \\
\cmidrule(lr){2-3} \cmidrule(lr){4-5}
Noise rate & No-Repair & Repair (avg) & No-Repair & Repair (avg) \\
\midrule
0.00 & \textbf{0.600} & 0.540 & \textbf{1.000} & 0.921 \\
0.15 & \textbf{0.555} & 0.413 & \textbf{0.946} & 0.896 \\
0.30 & \textbf{0.515} & 0.465 & \textbf{0.878} & 0.866 \\
0.45 & \textbf{0.455} & 0.433 & 0.792 & \textbf{0.832} \\
0.60 & 0.340 & \textbf{0.380} & 0.690 & \textbf{0.784} \\
\bottomrule
\end{tabular}
\caption{Regime structure of repair (Probe 2, $N=200$ dialogues per cell). On retrieval, no-repair wins at every noise rate up to $0.45$; repair only catches up at $0.60$. On meaning preservation, the curves \textit{cross} between $0.30$ and $0.45$, with repair becoming substantively better at high noise. The two metrics are answering different questions: retrieval requires preserving discriminative detail; cos-to-clean requires preserving gist meaning.}
\label{tab:regime}
\end{table}

\begin{figure}[h!]
\centering
\includegraphics[width=\textwidth]{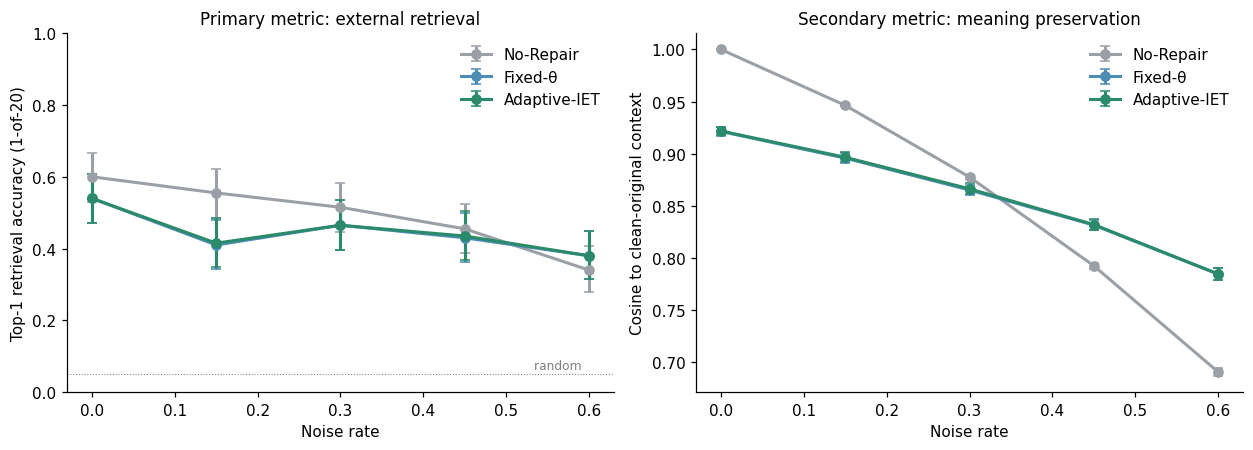}
\caption{Dual-metric headline from Probe 2. Left: 1-of-20 retrieval accuracy by noise rate. No-repair dominates at low and moderate noise; the repair conditions only become competitive at the highest noise rate, and the gain is small and not statistically significant. Right: cosine similarity to clean-original context. The no-repair line drops faster than the repair line as noise increases; the curves cross between noise rates $0.30$ and $0.45$. The two panels answer different questions, and the answer to ``does repair help?'' depends on which question is being asked.}
\label{fig:dualmetric}
\end{figure}

The clearest way to state the finding is this. \textbf{Repair preserves gist meaning under high noise but never recovers discriminative detail, regardless of controller.} This is visible in Figure~\ref{fig:dualmetric}: on the secondary metric, the no-repair line drops faster than the repair line as noise increases, and the curves cross between noise rates $0.30$ and $0.45$. On the primary metric, the no-repair line dominates from $0$ to $0.45$ and the repair lines only catch up at $0.60$ --- and even there the gain on retrieval is small and not statistically significant. Repair-based controllers, in this experimental setup, trade discriminative fidelity for gist fidelity; the trade only pays off when noise is high enough that the no-repair baseline is also losing discriminative fidelity at a comparable rate.

\subsection{What We Take From This}

We take three lessons from these probes, in increasing order of generality.

First, a methodological point. \textit{Pre-flight signal validation}, in which a candidate signal is tested against ground-truth labels before being fed to a controller, is a small discipline that significantly clarifies what subsequent experiments are testing. In Probe 1, the absence of this step meant we spent compute and analysis on a controller built on a signal that could not, in principle, do what we were asking of it. In Probe 2, the same step revealed that the redesigned signal was strong (AUC $0.828$), which sharpened the interpretation of the subsequent null result on adaptive-vs-fixed: the controllers were operating on a real signal, and still the IET update produced no measurable advantage. We would suggest this discipline as a standard for empathy-related and repair-related controllers.

Second, the IET update rule \textit{as currently formalised} did not, in either probe, produce measurable gain over a well-tuned fixed threshold. We do not claim this refutes the framing; we claim it falsifies a particular operationalisation. The variance-based update is gentle by construction --- it treats high recent variance as evidence to widen the band, low variance as evidence to tighten it --- and on dialogue-embedding signals at the scale we tested, the variance signal does not move enough to drive substantive controller differentiation. Whether richer signals (pragmatic, multimodal, discourse-level) or different update functionals (e.g., level-based rather than variance-based) would change this is a real research question for future work.

Third, the regime-structured behaviour of repair is, we think, the more important finding for the field. \citet{dong2023revisit} and \citet{huang2026pardon} test dialogue robustness and conversational repair, but treat them as essentially monolithic: a model is robust or it is not, repair is appropriate or it is not. Our results suggest a finer picture. Whether repair-based controllers help \textit{at all} depends jointly on the noise regime and on which aspect of preservation one cares about. This is consistent with \citet{wu2025hidden}'s recent finding that \textit{partial} alignment can outperform forced consensus in volatile environments, and it generalises that observation: not just consensus, but \textit{repair itself} has a regime structure that current evaluation methodology averages over.

The observed trade-off between gist preservation and discriminative detail is not a failure of empathic regulation, but a consequence of it: maintaining interpretive contact under noise requires sacrificing detail to preserve a shared band of meaning. We read the trade-off itself as substantively informative for the IET framing. Read this way, the regime structure is not an inconvenient finding about repair controllers — it is what we would expect a system optimised for interpretive homeostasis to do.

\section{Future Directions}

Four lines of work follow from the analysis above.

\paragraph{Multimodal and pragmatic operationalisation.}
Sentence-embedding distance, even with a well-chosen reference frame, is a coarse measure of interpretive divergence. Operationalisations that combine pragmatic features (clarification requests, hedging, repair initiation), prosodic and timing features (latency, overlap, pause structure), and physiological signals where available are likely to provide a more discriminating $\Delta_t$. Cross-modal IET implementations would also test whether the framing's predictions hold in modalities where the synchronic ``recognition'' picture is most entrenched.

\paragraph{Empirical psychology of tolerance.}
The IET lens treats $\theta_t$ as a regulatable individual and dyadic variable. This is a testable claim about human dialogue. Longitudinal studies of conversational dyads could examine whether more adaptive tolerance --- in the IET sense --- predicts relationship satisfaction, conflict resilience, and prosocial behaviour. Developmental studies could trace how interpretive tolerance emerges from early attachment dynamics into adult social cognition. Cross-cultural studies could test the variation in $\sigma^{*}$ that we have argued is a normative rather than descriptive parameter.

\paragraph{Regime-aware evaluation methodology.}
The Probe 2 result suggests a concrete methodological reform. Empathic-AI and dialogue-repair benchmarks should report performance \textit{across} a noise or perturbation grid, not at a single condition or as an average. They should also include at least two metrics with different sensitivities --- one harsh (retrieval, discrimination) and one forgiving (similarity, gist). Single-condition or single-metric evaluation is currently masking the kind of regime structure our results made visible. We see this as a low-cost change with high information value.

\paragraph{Long-horizon human--AI dyads and the ethics of relational design.}
Finally, the most demanding test of the framing is whether it predicts anything specific about extended human--AI interaction. Do users in long-term interaction with conversational agents show measurable adaptation of their own interpretive tolerance? Do agents whose tolerance bands are explicit and user-adjustable produce different relational outcomes than those whose tolerance is fixed and opaque? Do they preserve, as \citet{shukla2026relationship} argue they should, a user's relational ecology rather than collapsing it into a one-to-one bond with the system? These are questions that current evaluation frameworks for conversational AI do not even ask, and we suggest they should.

\section{Conclusion}

The value of this work is not in demonstrating a superior controller, but in reframing what should count as success in empathic interaction --- and in showing, through two careful and falsifiable attempts at operationalisation, where that reframing meets and resists empirical translation.

Taken together, our findings suggest that the central challenge in modelling empathy for extended interaction is not achieving alignment, but regulating divergence across time. The regime-structured nature of repair we surfaced in Probe 2 highlights that the value of intervention depends on the underlying noise conditions and on the aspect of meaning one seeks to preserve: repair preserves gist meaning under high noise but never recovers discriminative detail, regardless of controller. In this light, IET is best understood not as a fixed solution, but as a lens for rethinking empathy as a dynamic process of maintaining intelligibility across difference.

The picture that emerges differs in a small but consequential way from the dominant framing of empathic AI. Empathy, on this account, is not the disappearance of difference between minds. It is the work of staying in contact across difference --- of inhabiting the moving threshold where two interpretive trajectories remain mutually intelligible without becoming the same. If prediction is the language of the brain, empathy is the rhythm of coexistence: the capacity to anticipate divergence, and to choose, again, to remain in meaning.

\section*{Statements and Declarations}
\textbf{Funding.} No external funding supported this work.\\
\textbf{Competing Interests.} The author declares no competing interests.\\
\textbf{Data Availability.} The computational probes use publicly available data (DailyDialog) and standard libraries (sentence-transformers); the analysis notebooks and pre-committed hyperparameter values are available from the author on request and will be deposited in a public repository on acceptance.\\
\textbf{AI Use.} Drafting and editorial revision of this manuscript involved use of an AI writing assistant; all conceptual content, formalism, experimental design, and argumentation are the author's, and the author takes full responsibility for the final text.

\bibliographystyle{plainnat}
\bibliography{references}

\end{document}